\documentclass[twocolumn,showpacs,preprintnumbers,showkeys,amsmath,amssymb]{revtex4}
\usepackage{graphicx}
\usepackage{dcolumn}
\usepackage{bm}
\begin{document}
\title{3D Lattice-Boltzmann Model for Magnetic Reconnection}
\author{M. Mendoza}
 \email{mmendozaj@unal.edu.co}
\affiliation{
Simulation of Physical Systems Group, Universidad Nacional de Colombia, Departamento de Fisica,\\
Crr 30 \# 45-03, Ed. 404, Of. 348, Bogot\'a D.C., Colombia\\
 }
\author{J. D. Munoz}
 \email{jdmunozc@unal.edu.co}
\affiliation{
Simulation of Physical Systems Group, Universidad Nacional de Colombia, Departamento de Fisica,\\
Crr 30 \# 45-03, Ed. 404, Of. 348, Bogot\'a D.C., Colombia\\
}

\date{\today}
\begin{abstract}
  In this paper we develop a 3D Lattice-Boltzmann model that recovers
  in the continuous limit  the two-fluids theory for plasmas, and
  consecuently includes the generalizated Ohm's law. 
  The model reproduces the magnetic reconnection
  process just by given the right initial equlibrium conditions in the
  magnetotail, without any assumption on the resistivity in the diffusive
  region. In this model, the plasma is handled like two fluids with an
  interaction term, each one with distribution functions associated to a
  cubic lattice with 19 velocities (D3Q19). The electromagnetic fields
  are considered like a third fluid with an external force on a
  cubic lattice with 13 velocities (D3Q13). The model can simulate
  either viscous fluids in the incompressible limit or non-viscous
  compressible fluids, and sucessfully reproduces both the Hartmann
  flow and the magnetic reconnection in the magnetotail. The
  reconnection rate obtained with this model is $R$$=$$0.109$, which
  is in excellent agreement with the observations.
\end{abstract}



\maketitle
\section{Introduction}
The magnetic reconnection is one of the most interesting phenomenon of
plasma physics. This process quickly transforms the magnetic energy into termic
and kinetic energies of the plasma. It is mostly observed  inside of
astrophysical plasmas, such as solar flares (where it contributes to
the plasma heating), and in the terrestrial magnetosphere, where it
support the income flux of plasma and electromagnetic energy. 

The magnetic
reconnection requires the existence of a diffusive region, where 
dissipative electric fields change the magnetic field topology.   
The first models were independently formulated by Sweet \cite{n0}, in 1958,
and Parker \cite{n1}, in 1957. 
They suggested that the magnetic reconnection is a steady-state
resistive process that  occurs in the vicinity of a neutral line. 
This model reduces the phenomenon to a boundary condition problem and
can explain the magnetic field reconnection. However, it has
some problems when compared with experimental observations 
(i.e. a very slow reconnection rate), and it leaves unexplained the
origin of the high-resistive region. 
In 1964, Petschek  \cite{n2} proposed the first model
for fast reconnection rates. He included a much smaller diffusion
region than the Sweet-Parker model, but he suggested that the rest of the boundary
layer region should consist of slow shock waves that accelerate the
plasma up to the Alfven velocity. Nevertheless, the origin of the
diffusive region remains unexplained. 

At present, the nature of this phenomenon has been studying by using 
kinetic theory and considering collisionless plasmas, since this is a common
property of astrophysical plasmas. One of the developments of the
kinetic theory is the generalized Ohm's law, where some extra
terms explain the existence of a dissipative electric field. The introduction
of these extra terms in resistive magnetohydrodynamics is called
MHD-Hall \cite{n3}.
A useful approximation of the kinetic theory consists of modelling the
plasma like two fluids (one electronic and one ionic), which have
independent momentum, mass conservation and state equations, plus an
interaction term in the momentum equation \cite{n3}. This treatment, in
the one-fluid limit, introduces in a natural way the extra terms of
the generalized Ohm's law.  However, the equations involved by this
treatment are complex and it is difficult to find an analytic solution
for any problem. 
 
For this reason, most plasma processes are studied by numerical methods.  
One of the numerical methods for simulating fluids is Lattice
Boltzmann (LB) \cite{n5}, which was developed from lattice-gas automata. 
Lattice Boltzmann simulations are performed on regular grids of many
cells and a small number of velocity vectors per cell, each one
associated to a density distribution function, which evolve and spread
together to the neighbohr cells  according to the collisional
Boltzmann equation.
The first LB model for studying plasmas reproduces the resistive
magnetohydrodynamic equations and was developed by Chen \cite{n6,n7} as an
extension of the Lattice-Gas model developed by 
Chen and Matthaeus\cite{n8} and Chen, Matthaeus and Klein \cite{n9}. This LB model uses
37 velocity vectors per cell on a square lattice and is developed for
two dimensions.    
Thereafter, Martinez, Chen and Matthaeuss \cite{n10} decreased the number
of velocity vectors from 37 to 13, which made easier a future 3D
extension. 
One of the first  LB models for magnetohydrodynamics in 3D was
developed by Bryan R. Osborn in his master thesis \cite{n11}. He used 19
vectors on a cubic lattice for the fluid, plus 7 vectors for the
magnetic field, which makes a total number of 26 vectors per cell.
By following a different path, Fogaccia, Benzi and Romanelli \cite{n12}
introduced a 3D LB model for simulating turbulent plasmas in the
electrostatic limit. All these models reproduce the resistive
magnetohydrodynamc equations for a single fluid.
 
In this paper, we introduce a 3D Lattice-Boltzmann model that
recovers the plasma equations in the two-fluids theory. In this way,
the model is able to reproduce magnetic reconnection, without the
{\it a priori} introduction of a resistive region. Moreover, it is
able to reproduce the fluid state-equation with a general polytropic
coefficient. The model uses 39 vectors per cell and 63 probability
density functions (19 for each fluid, 25 for the electrical and magnetic
fields). In section \ref{LBmodel} we describe the model, with the evolution
rules and the equilibrium expressions involved for the 63 density functions,
plus the way to compute the electric, magnetic and velocity fields. The
Chapman-Enskog expansion showing how these rules recover the two-fluids
magnetohydrodynamic equations is developed in Appendix \ref{ChapmanEnskog}. In order to
validate the model, we simulate the 2D Hartmann's flow in section
\ref{Hartmann}, and, finally, the magnetic reconnection for a magnetotail
equilibrium configuration in section \ref{Magnetotail}. The main results and
conclusions are summarized in section \ref{conclusion}.  

\section{3D Lattice-Boltzmann Model for a Two-Fluids Plasma}
\label{LBmodel}
In a simple Lattice-Boltzmann model \cite{n5}, the $D$-dimensional space
is divided into a regular grid of cells. Each cell 
has $Q$ vectors $\vec{v}_i$ that links itself with its
neighbors, and each vector is associated to a distribution function
$f_i$. 
The distribution function evolves at time steps $\delta t$ according to the
Boltzmann equation,
\begin{equation}{\label{boltzeq}}
  f_i(\vec{x}+\vec{v}_i \delta t,t+\delta t)-f_i(\vec{x},t)= \Omega_i(\vec{x},t) \quad ,
\end{equation}
where $\Omega_i(\vec{x},t)$ is a collision term, which is usually
taken as a time relaxation to some equilibrium density, $f_i^{\rm
  eq}$. This is known as the the Bhatnagar-Gross-Krook (BGK) operator \cite{n13}, 
\begin{equation}{\label{collop}}
  \Omega_i(\vec{x},t)=-\frac{1}{\tau}(f_i(\vec{x},t)-f_i^{\rm
  eq}(\vec{x},t)) \quad ,
\end{equation}
where $\tau$ is the relaxation time and $f_i^{\rm eq}(\vec{x},t)$ is the equilibrium
function. The equilibrium function is chosen in such a way, that (in
the continuum limit) the model simulates the actual physics of the system.
\begin{figure}
  \centering
  \includegraphics[scale=0.5]{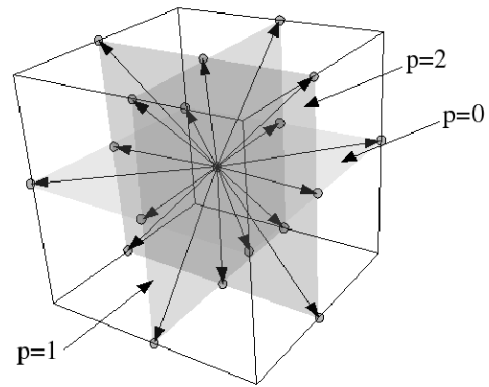}
  \caption{Cubic Lattice D3Q19 for modelling the electronic and ionic fluids.
  The arrows represent the velocity vectors $\vec{v}_{i}^{p}$ and $p$ indicates the
  plane of location.}\label{d3q19}
\end{figure}
\begin{figure}
  \centering
  \includegraphics[scale=0.5]{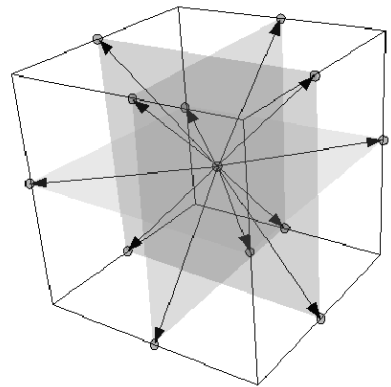}
  \caption{Cubic Lattice D3Q13 for modelling the electric field.
  The arrows represent the electric vectors $\vec{e}_{ij}^{p}$.}\label{d3q13}
\end{figure}
\begin{figure}
  \centering
  \includegraphics[scale=0.5]{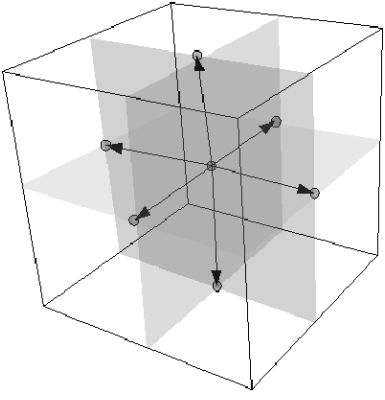}
  \caption{Cubic Lattice D3Q7 for simulating the magnetic field,
  the arrows indicate the magnetic vectors $\vec{b}_{ij}^{p}$.}\label{d3q7}
\end{figure}
\begin{figure}
  \centering
  \includegraphics[scale=0.5]{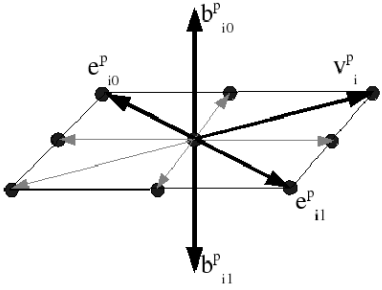}
  \caption{Index relationship between the velocity vectors and the electric
  and magnetic vectors.}\label{conexion} 
\end{figure}

For our 3D model, we use a cubic regular grid, with lattice constant
$\delta x$$=$$\sqrt{2} c\delta t$ and $c$ is the light speed
($c$$\simeq$$3\times10^8 m/s$). There are 19 velocity vectors for 
the electronic and ionic fluids (figure \ref{d3q19}), 13 vectors for the
electric field (figure \ref{d3q13}) and  7 vectors for the
magnetic field (figure \ref{d3q7}). The velocity vectors are denoted 
by $\vec{v}_i^p$, where $i=1,2,3,4,5,6$ indicates the direction and
$p=0,1,2$ indicates the plane of location. Their components are  
\begin{subequations}{\label{coord}}
  \begin{equation}
    \vec{v}_i^0=c\sqrt{2}(\cos((2i-1)\pi/4),\sin((2i-1)\pi/4),0) \quad ,
  \end{equation}
  \begin{equation}
    \vec{v}_i^1=c\sqrt{2}(\cos((2i-1)\pi/4),0,\sin((2i-1)\pi/4)) \quad ,
  \end{equation}  
  \begin{equation} 
    \vec{v}_i^2=c\sqrt{2}(0,\cos((2i-1)\pi/4),\sin((2i-1)\pi/4)) \quad ,
  \end{equation}
\end{subequations}
for $i<5$, and 
\begin{subequations}{\label{coord2}}
  \begin{equation}
    \vec{v}_i^0=c\sqrt{2} ((-1)^i,0,0) \quad ,
  \end{equation}
  \begin{equation}
    \vec{v}_i^1=c\sqrt{2}(0,(-1)^i,0) \quad ,
  \end{equation}  
  \begin{equation} 
    \vec{v}_i^2=c\sqrt{2}(0,0,(-1)^i) \quad ,
  \end{equation}
\end{subequations}
for $i\ge 5$. This makes 18 vectors. The missing one is the rest vector $\vec{v}_0$,
with componets $(0,0,0)$.

 The set of 13 electric field vectors, $\vec{e}_{ij}^p$, and 7
magnetic field vectors, $\vec{b}_{ij}^p$ are related with the velocity
vectors as follows: 
\begin{equation}
  \vec{e}_{i0}^p=\frac{1}{2}\vec{v}_{(i-1)mod 4}^p 
  \quad , \quad
  \vec{e}_{i1}^p=\frac{1}{2}\vec{v}_{(i+1)mod 4}^p \quad ,
\end{equation}
and
\begin{equation}
  \vec{b}_{ij}^p=\frac{1}{2c^2} \vec{v}_i^p \times \vec{e}_{ij}^p \quad ,
\end{equation}
where the index $i$ takes the values $i$$=$$1,2,3,4$.

The distribution functions that describe the fluids, denoted by
$f_{i}^{p(s)}$ and $f_0^{(s)}$, propagate with each velocity vector
$\vec{v}_i^p$ and with the rest vector $\vec{v}_0$, respectively, and
uses these vectors to compute the velocity fields for each fluid. 
Here, the index $s$ distinguishs between electronic ($s$$=$$0$) and
ionic ($s$$=$$1$) fluids. 
Similarly, the distribution functions associated for the electromagnetic field
are denoted by $f_{ij}^{p(2)}$ and $f_0^{(2)}$. They also propagate in the
direction of the velocity vectors $\vec{v}_i^p$ and $\vec{v}_0$, but
they use the electric and magnetic field vectors to compute those
fields. Summarizing, The macroscopic variables are computed as follows: 
\begin{subequations}{\label{macros}}
  \begin{equation}
    \rho_s=f_0^s + \sum_{i,p} f_{i}^{p(s)} \quad ,
  \end{equation}
  \begin{equation}
    \rho_s \vec{V_s}=\sum_{i,p} f_{i}^{p(s)} \vec{v}_i^p \quad ,
  \end{equation}
  \begin{equation}
    \vec{E}=\sum_{i,j,p} f_{ij}^{p(2)} \vec{e}_{ij}^p \quad ,
  \end{equation}
  \begin{equation}
    \vec{B}=\sum_{i,j,p} f_{ij}^{p(2)} \vec{b}_{ij}^p \quad ,
  \end{equation}
  \begin{equation}
    \vec{J}=\sum_s \frac{q_s}{m_s} \rho_s \vec{V}_s \quad ,  
  \end{equation}
  \begin{equation}
    \rho_c=\sum_s \frac{q_s}{m_s} \quad ,  
  \end{equation}
\end{subequations}
where $\rho_s$ and $\vec{V}_s$ are the density and velocity of each
fluid, and $m_s$ and $q_s$ are its particle mass and charge (here,
$s$$=$$0$ represents electrons and $s$$=$$1$ represents ions, as before).
In addition, $\vec{E}$ and $\vec{B}$ are the electric and magnetic
fields, $\vec{J}$ is the total current density and $\rho_c$ is the
total charge density.

For their evolution, we follow the proponsal of J.M. Buick and C.A. Greated
for the lattice Boltzmann equations \cite{n14},  
\begin{eqnarray}{\label{lbe0}}
  \begin{aligned}
    f_{i}^{p(s)}(\vec{x}+\vec{v}_i^p \delta t,t+\delta
    t)-f_{i}^{p(s)}(\vec{x},t)=& \\
    \Omega_{i}^{p(s)}(\vec{x},t)+&\frac{\kappa_s \delta t}{20c^2}(\vec{v}_i^p\cdot 
    \vec{F}^{(s)}) ,
  \end{aligned}
\end{eqnarray}
\begin{eqnarray}{\label{lbe2}}
  \begin{aligned}
  f_{ij}^{p(2)}(\vec{x}+\vec{v}_i^p \delta t,t+\delta
  t)-f_{ij}^{p(2)}(\vec{x},t)=&  \\
  \Omega_{ij}^{p(2)}(\vec{x},t)-&\frac{\kappa_2 \mu_0 \delta t}{8}(\vec{e}_{ij}^p \cdot
  \vec{J'}) ,
  \end{aligned}
\end{eqnarray}
\begin{eqnarray}{\label{lbe3}}
  f_{0}^{(K)}(\vec{x},t+\delta t)-f_{0}^{(K)}(\vec{x},t)= \Omega_{0}^{(K)}(\vec{x},t) \quad ,
\end{eqnarray}
where $K=0,1,2$. The force vectors $\vec{F}^{(s)}$ in Eq.\eqref{lbe0} are 
\begin{eqnarray}{\label{force}}
  \begin{aligned}
    \vec{F}^{(s)}=&\frac{q_s}{m_s}\rho_s(\vec{E}+\vec{V}_s \times
    \vec{B}) \\ &-\nu \rho_s(\vec{V}_s-\vec{V}_{(s+1)mod 2})+\vec{F}^{(s)}_0 \quad ,
  \end{aligned}
\end{eqnarray}
where $\nu$ is the collision frequency of the plasma, $\vec{F}^{(s)}_0$ is any
external force (for instance, a gravitational force) and the equilibrium
density current vector $\vec{J'}$ in Eq. \eqref{lbe2} is defined by
\begin{equation}{\label{currentequil}}
    \vec{J'}=\sum_s \frac{q_s}{m_s} \rho_s \biggl (\vec{V}_s +\frac{\lambda_s \tau_s \delta t
  \vec{F}^{(s)}}{\rho_s} \biggr ) \quad .  
\end{equation}

The collision terms $\Omega_{ij}^{p(K)}$ and $\Omega_{0}^{(K)}$ are given by
\begin{subequations}
  \begin{equation}
    \Omega_{i}^{p(s)}=-\frac{1}{\tau_s}(f_{i}^{p(s)}(\vec{x},t)-f_{i}^{p(s)\rm
    eq}(\vec{x},t))\quad ,
  \end{equation}
  \begin{equation}
    \Omega_{ij}^{p(2)}=-\frac{1}{\tau_2}(f_{ij}^{p(2)}(\vec{x},t)-f_{ij}^{p(2)\rm
    eq}(\vec{x},t))\quad ,
  \end{equation}
  \begin{equation}
    \Omega_{0}^{(K)}=-\frac{1}{\tau_K}(f_{0}^{(K)}(\vec{x},t)-f_{0}^{(K)\rm
    eq}(\vec{x},t))\quad ,
  \end{equation}
\end{subequations}
where $\tau_K$ is the relaxation time,
$\kappa_{K}$$=$$\frac{2\tau_K-1}{2 \tau_K}$ and $\lambda_{s}$$=$$\frac{1}{2
  \tau_s}$. 

The equilibrium functions for the fluids, $f_{i}^{p(s)\rm eq}$ and
$f_{0}^{(s)\rm eq}$ are 
\begin{subequations}{\label{equilf}}
\begin{eqnarray}
  \begin{aligned}
    f_{i}^{p(s)\rm eq}(\vec{x},t)=&\omega_i \rho_s  \biggl[ 3 \xi_s
    \rho_s^{\gamma-1}+3(\vec{v}_i^p \cdot
    \vec{V'}_s) \biggr.  \\ & \biggl. +\frac{9}{4c^2}(\vec{v}_i^p \cdot  \vec{V'}_s)^2 -
    \frac{3}{2}(\vec{V'}_s^2) \biggr]\quad ,
  \end{aligned}
\end{eqnarray}
\begin{equation}
  f_{0}^{p(s)\rm eq}(\vec{x},t)=6 \rho_s c^2 \biggl[1-\frac{1}{4c^2}(4\xi_s
  \rho_s^{\gamma-1}+\vec{V'}_s^2) \biggr]\quad ,
\end{equation}
\end{subequations}
where the weights $w_i$ are $w_0=\frac{1}{6c^2}$, $w_{1,2,3,4}=\frac{1}{72c^2}$,
$w_{5,6}=\frac{1}{36c^2}$. In addition, $\xi_s$ is a constant that is
fixed by the initial fluid temperature and density by means of the
ideal gas law,
\begin{equation}{\label{Temperature}}
  \xi_s=\rho_{s(t=0)}^{1-\gamma}\frac{k}{m_s}T_{s(t=0)}\quad ,
\end{equation}
with polytropic index $\gamma$, and $k$ is the Boltzmann constant.
The equilibrium velocity $\vec{V'}_s$ is defined by 
\begin{equation}{\label{expandV}}
  \vec{V'}_{s}=\vec{V}_{s}+\frac{\lambda_s \tau_s \delta t
  \vec{F}^{(s)}}{\rho_s} \quad .
\end{equation}

For the electromagnetic field ($K=2$), we have
\begin{subequations}{\label{equilc}}
\begin{equation}
  f_{ij}^{p(2) \rm eq}(\vec{x},t)=\frac{1}{8c^2}\vec{E'} \cdot
  \vec{e}_{ij}^{p}+\frac{1}{8}\vec{B} \cdot \vec{b}_{ij}^{p} \quad ,
\end{equation}
\begin{equation}
  f_{0}^{(2)\rm eq}(\vec{x},t)=0 \quad ,
\end{equation}
\end{subequations}
where the equilibrium electric field is
\begin{subequations}{\label{expandE}}
\begin{equation}
  \vec{E'}=\vec{E}-(\mu_0 c^2 \lambda_2 \tau_2 \delta t)\vec{J'} \quad , 
\end{equation}
\end{subequations}
and $\lambda_{2}$$=$$\frac{1}{2 \tau_2}$, as before.

The proof that this lattice Boltzmann model, via a Chapman-Enskog
expansion, recovers the equations of the two-fluids theory for a plasma
composed by electrons and ions is shown in Appendix \ref{ChapmanEnskog}.
The model let us to consider either compressible and non-viscous
fluids or incompressible and viscous 
fluids. The first ones are governed by the continuity equation
\begin{eqnarray}{\label{ec}}
  \vec{\nabla} \cdot (\rho_s {\vec{V}'}_s)+\frac{\partial \rho_s}{\partial
  t}=0 \quad ,
\end{eqnarray}
the Navier-Stokes equation,
\begin{eqnarray}{\label{Nefc}}
  \begin{aligned}
    \rho_s \biggl (\frac{\partial \vec{V'}_S}{\partial t} +& (\vec{V'}\cdot
      \vec{\nabla}) \vec{V'}_s \biggr)=\\-&\vec{\nabla} P_s+\frac{q_s}{m_s}\rho_s(\vec{E}+\vec{V'}_s \times
    \vec{B}) \\ -&\nu
    \rho_s(\vec{V'}_s-\vec{V'}_{(s+1)mod 2})+\vec{F}_{0} \quad .
  \end{aligned}
\end{eqnarray}
the state equation,
\begin{eqnarray}{\label{es}}
  P_s=\xi_s \rho_s^{\gamma} \quad ,
\end{eqnarray}
where $P_s$ is the fluid pressure, and the Maxwell equations. The second ones are governed by the state
equation \eqref{es}, Maxwell equations, the continuity equation
\begin{eqnarray}{\label{ec}}
  \vec{\nabla} \cdot  \vec{V}'_s =0 \quad 
\end{eqnarray}
and the Navier-Stokes equation for an incompressible and viscous fluid,
\begin{eqnarray}{\label{Nefi}}
  \begin{aligned}
    \rho_s \biggl (\frac{\partial \vec{V'}_s}{\partial t} &+ (\vec{V'}\cdot
      \vec{\nabla}) \vec{V'}_s \biggr)=\\&-\vec{\nabla} P_s+\frac{q_s}{m_s}\rho_s(\vec{E}+\vec{V'}_s \times
    \vec{B}) \\ &-\nu
    \rho_s(\vec{V'}_s-\vec{V'}_{(s+1)mod 2})\\ &+ \vec{F}_{0} +\eta_s
    \rho_s \vec{\nabla}^2 \vec{V'}_{s}\quad .
  \end{aligned}
\end{eqnarray}
where the kinematic viscosity is $\eta_s$$=$$\frac{2}{3}(\tau_s-1/2)c^2 \delta t$.

\section{Simulation of a 2D Hartmann Flow}
\label{Hartmann}
In the MHD limit, the two-fluid theory becomes the MHD (one fluid) theory, which is
represented by the following equations: the continuity of mass,
\begin{eqnarray}{\label{cMHD}}
  \vec{\nabla} \cdot (\rho \vec{V})+\frac{\partial \rho}{\partial t}=0 \quad ,
\end{eqnarray}
the Navier-Stokes equation,
\begin{eqnarray}{\label{vMHD}}
  \rho \biggl( \frac{\partial}{\partial t}+\vec{V} \cdot \vec{\nabla}
  \biggr)\vec{V}=-\vec{\nabla} P + \vec{J}\times \vec{B} + \eta \vec{\nabla}^2
  \vec{V} + \vec{F}_{0} ,
\end{eqnarray}
the magnetic field equation,
\begin{eqnarray}{\label{bMHD}}
  \frac{\partial \vec{B}}{\partial t}=\vec{\nabla} \times
  (\vec{V}\times\vec{B})+\eta_m \vec{\nabla}^2 \vec{B} \quad ,
\end{eqnarray}
and the state equation,
\begin{eqnarray}{\label{eMHD}}
  P=\xi_s \rho^\gamma \quad ,
\end{eqnarray}
where $\rho$ is the total mass density, $\vec{V}$ is the total
velocity field and $\eta_m$$=$$\frac{1}{\mu_0 \sigma_0}$ is the
magnetic viscosity.
 
For the Hartmann flow \cite{n15,n16}, we consider a fluid in isotermal equilibrium
($\gamma=1$) at low temperature (a small $\xi_s$ value),
incompressible and viscous. The fluid moves in the $x$ direction between
two walls at rest at $y$$=$$-L$ and $y$$=$$-L$.
There is a constant magnetic field in the $y$ direction, with
intensity $B_0$, and a constant external force $F$$=$$\rho g$ in the $x$
direction to drag the fluid \cite{n15}. 
So, the velocity and magnetic fields take the forms $\vec{V}$$=$$(V_x(y),0,0)$ and 
$\vec{B}=(B_x(y),B_0,0)$, respectively. By replacing these expressions in 
equations \eqref{vMHD} and \eqref{bMHD}, one finds the following solutions for the
velocity and magnetic fields \cite{n15}: 
\begin{subequations}{\label{solutionMHD}}
\begin{eqnarray}
  V_x(y)=\sqrt{\frac{\rho \eta_m}{\eta}} \frac{g L}{B_0}
  \cosh(H)\biggl[1-\frac{\cosh(Hy/L)}{\cosh(H)} \biggr] \quad ,
\end{eqnarray}
\begin{eqnarray}
  B_x(y)=\frac{\rho g L}{B_0}\biggl[\frac{\sinh(Hy/L)}{\sinh(H)}
  -\frac{y}{L} \biggr] \quad ,
\end{eqnarray}
\end{subequations}
where $H$$=$$\frac{B_0 L}{\sqrt{\rho \eta \eta_m}}$ is the Hartmann number and
$-L\leq y \leq L$. 

For the simulation, we use a single row of $80$ cells in the $y$
direction, with periodic boundary conditions in both $x$ and $z$
directions. The initial conditions for the density functions are
obtained from the equilibrium expressions \eqref{equilf} and \eqref{equilc} with the values
$\vec{V}_s$$=$$0$, $\rho_s$$=$$m_sn_s$, $\vec{E}$$=$$0$,
$\vec{B}$$=$$(0,B_0,0)$ and $\vec{F}^{(s)}_0$$=$$(\rho_sg,0,0)$.
In addition, the constant values are $\gamma$$=$$1$, $\xi_s$$=$$3\times
10^{-6}$, $\mu_0$$=$$1.0$, $c$$=$$1$, $\nu$$=$$100$, $\tau_s$$=$$1.0$,
$\tau_2$$=$$0.5$, $m_0$$=$$1.0 \times 10^{-19}$, $m_1$$=$$1820m_0$,
and $n_0$$=$$n_1$$=$$1.0 \times 10^{19}$ particles per unit volume.
For the $y$ direction, we assume as boundary conditions at the walls that the 
equilibrium density functions for the time evolution (Eq. \eqref{equilf} and \eqref{equilc}) are
always the same from the initial conditions (including
$\vec{V}_s$$=$$0$, i.e. non-conducting walls). The system evolves until a steady state is
reached. We ran simulations for Hartmann numbers $H$$=$$5$, $13$ and
$26$, and the magnetic field $B_0$ was chosen to obtain these Hartmann numbers.

\begin{figure}
  \centering
  \includegraphics[scale=0.45]{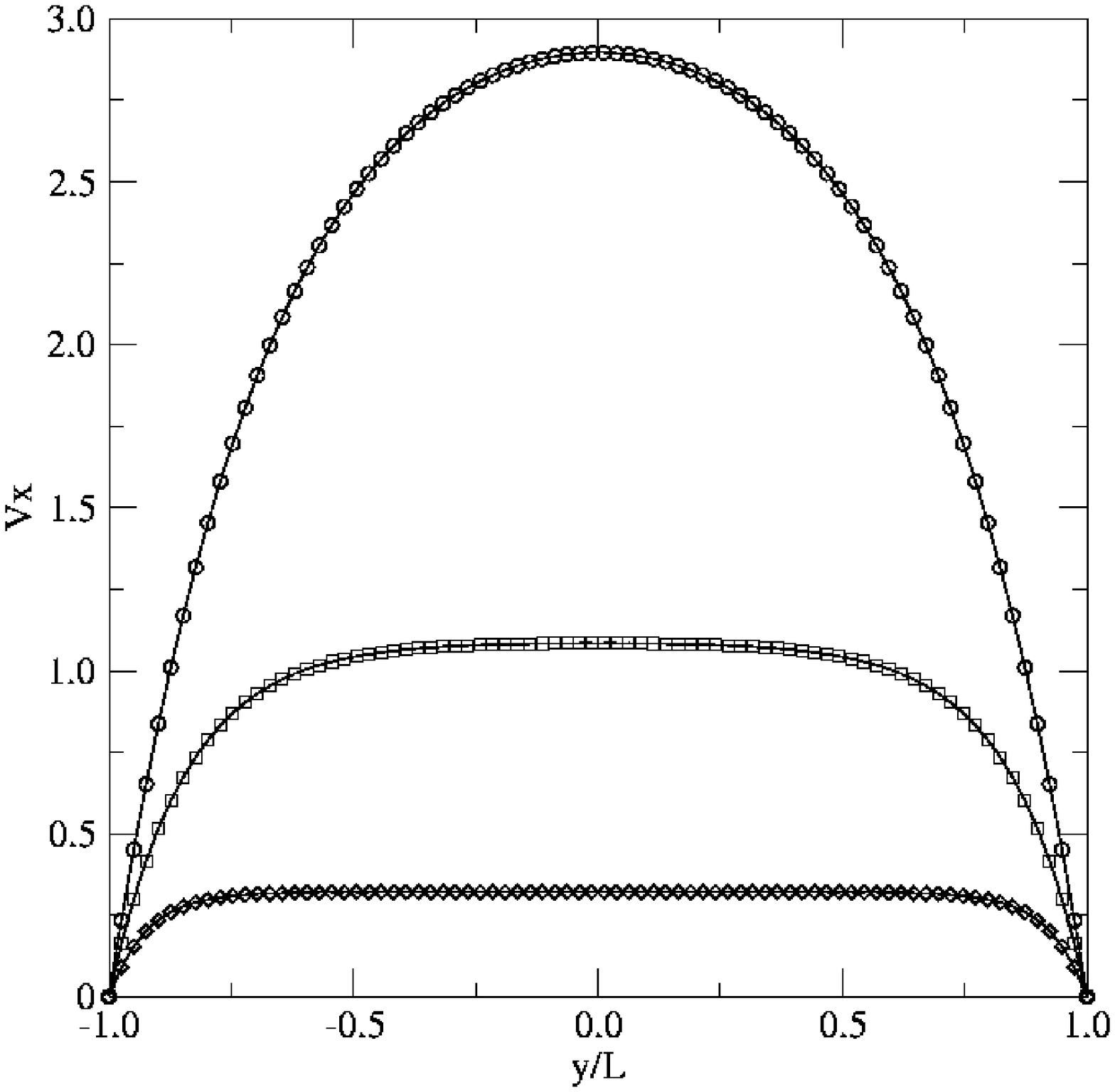}
  \caption{Velocity profile Vx vs. y/L for different Hartmann numbers: H=6.0
  (circles), H=13.0 (squares) and H=26.0 (diamonds). The solid lines are the
  analytical results.}\label{velocidadH}
\end{figure}

\begin{figure}
  \centering
  \includegraphics[scale=0.45]{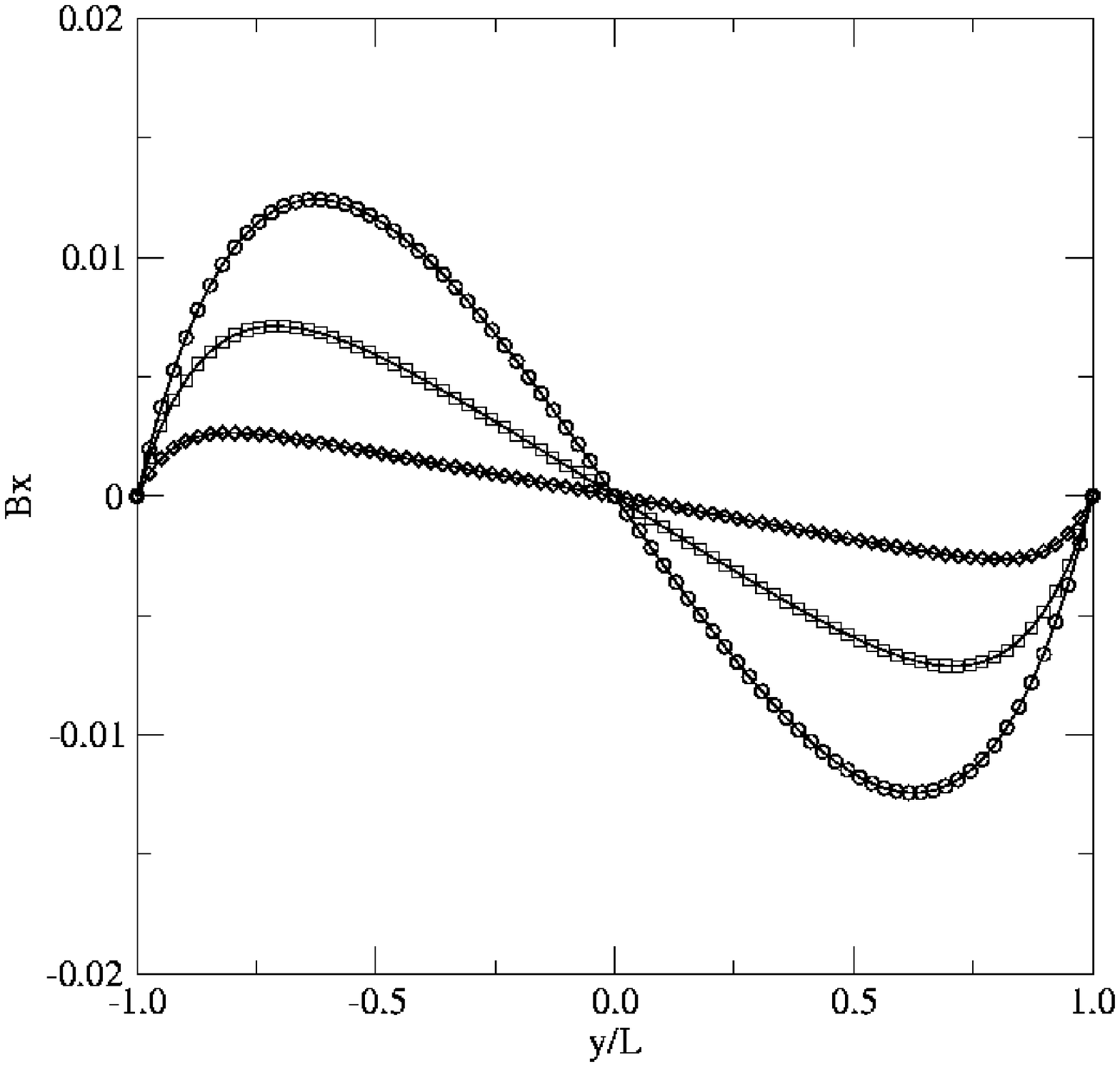}
  \caption{Magnetic field intensity Bx vs. y/L for different Hartmann numbers: H=6.0
  (circles), H=13.0 (squares) and H=26.0 (diamonds). The solid lines are the
  analytical results.}\label{campoH}
\end{figure}

Figure \ref{velocidadH} shows the velocity profiles and figure \ref{campoH} shows the
magnetic field profiles for the three cases. The solid lines are the
analytic solutions  (Eq.\eqref{solutionMHD}). The simulation results
are in excellent agreement with the analytical solutions. This result
say us that (at least for the MHD limit) our LB models works properly.  

\section{Application to Magnetic Reconnection}
\label{Magnetotail}
\subsection{Dynamics of the magnetic reconnection process}
In order to simulate the magnetic reconnection in the magnetotail,
we chose the initial equilibrium condition proposed by Harris
\cite{n17,n18} for the current sheet, plus a magnetic dipole field, ortogonal
to the sheet. For this simulation we assume that the fluids are non-viscous
and compressible.

The current sheet lies on the x-y plane, and its magnetic field is
described by the vector potential  $\vec{A}$$=$$(0,A_y,0)$, with
\begin{eqnarray}{\label{potential}}
  A_y(x,z)=L B_0 \ln{\cosh[v(x)(z/L)]/v(x)} \quad ,
\end{eqnarray}
where the effective thickness of the current sheet is given by $L/v(x)$, and
the asymptotic strength, $B_0$, is the value of $B_x$ in the limit $z
\to \infty$, divided by $v(x)$. The function $v(x)$ is an arbitrary
slowly-varying function. We choose for $v(x)$ the quasi-parabolic function
proposed by \cite{n19,n20}, 
\begin{eqnarray}{\label{tcurrent}}
  v(x)=\exp(-\epsilon x/L) \quad ,
\end{eqnarray}
where the parameter $\epsilon$ is much smaller than one and determines the
strength of the z-component of the magnetic field. We took $\epsilon$$=$$0.1$
for the simulation. The initial density is the one proposed by Harris,
\begin{eqnarray}{\label{tdensity}}
  n_s(x,z)=n_b+n_c v^2(x)\cosh^{-2}[v(x)(z/L)] \quad ,
\end{eqnarray}
where $n_b$ is the background density and $n_b+n_c$ is the maximal
density. 

The magnetic dipole is set at position $x_0$ with momentum $M$ and
oriented in the $z$ direction. It generates a magnetic field given by 
\begin{equation}\label{dipole}
  \begin{aligned}
    B_x(x,z)=&\frac{3M(x-x_0)z}{((x-x_0)^2+z^2)^{\frac{5}{2}}} \quad , \\ \\
    B_y(x,z)=&0 \quad , \\ \\
    B_z(x,z)=&\frac{M(2z^2-(x-x_0)^2)}{((x-x_0)^2+z^2)^{\frac{5}{2}}}
    \quad .
  \end{aligned}
\end{equation}

The lattice constant $\delta x$ is chosen as one seventh of the ion
inertial length, $\delta x$$=$$\frac{1}{7} c/\omega_1$, where
$\omega_1$ is the ion plasma frequency,
$\omega_1$$=$$\sqrt{\frac{q_1^2 n_1}{\epsilon_0 m_1}}$, with
$n_1$$=$$10^5$ particles per cubic meter for the magnetotail \cite{n21} and $m_1$
the proton mass. That gives $\delta x$$\simeq$$103$km. Since the
current sheet in the magnetotail can be assumed around $3000$km width
\cite{n21,n22}, we chose $L$$=$$2 c/\omega_1$. For the position of the magnetic
dipole, we took $x_0$$=$$22.7 c/\omega_1$ and for the dipole momentum,
$M$$=$$3\times 10^{12}$. The grid is an array of $100$$\times$$100$ cells on
the x-z plane with periodic boundary conditions in the $y$ direction and free
boundary conditions for the fields in the other directions (each boundary cell
copies the density functions of its first neighbohr in ortogonal
direction to the boundary at each time step). Thus, the simulation
region is a square of $14.26 c/\omega_1$ length (around $10300$km).
For this simulation we took $m_0$$=$$m_1/100$
(i.e. an electron mass 20 times larger than the real one) in
order to obtain numerical stability, but it has been shown \cite{n25}
that this point does not qualitatively change the physical results. 
The temperature ratio is chosen to be $T_0/T_1$$=$$0.2$, acording to
observational results \cite{n23}. For this simulation, we
took $n_c$$=$$ 5 n_b$ and $n_b$$=$$0.17 n_1$.
\begin{figure}
  \centering
  \includegraphics[scale=1.0]{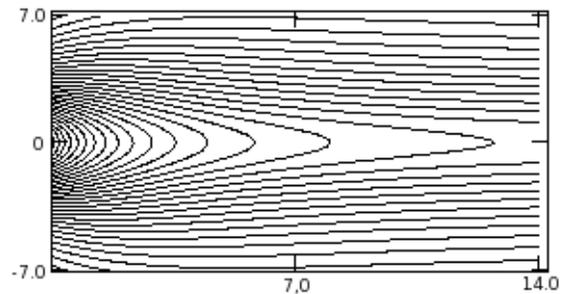}
  \caption{Magnetic field lines in the magnetic reconnection
    process at t=0 (initial conditions)}\label{tiempo0}
\end{figure}
\begin{figure}
  \centering
  \includegraphics[scale=1.0]{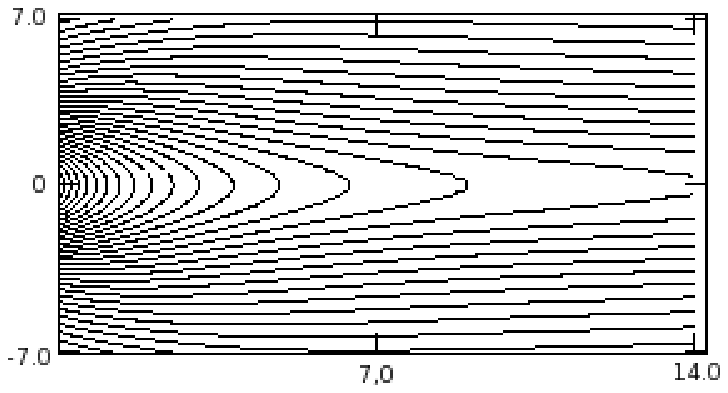}
  \caption{Evolution for the Magnetic field lines in the magnetic reconnection
  process, at $t=3/\Omega_1$}\label{tiempo3}
\end{figure}
\begin{figure}
  \centering
  \includegraphics[scale=1.0]{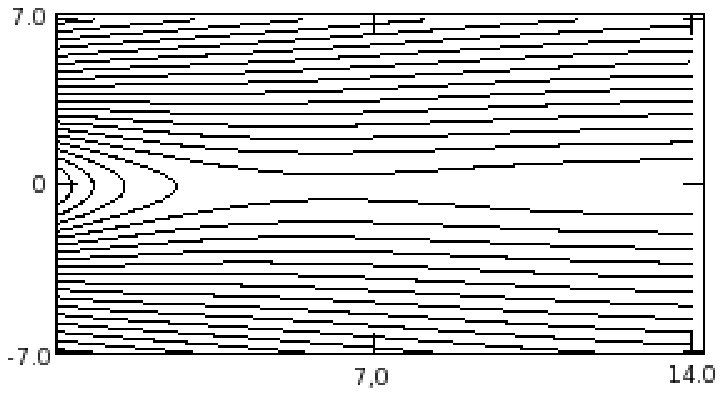}
  \caption{Evolution for the Magnetic field lines in the magnetic reconnection
  process, at $t=15/\Omega_1$}\label{tiempo15}
\end{figure}
\begin{figure}
  \centering
  \includegraphics[scale=1.0]{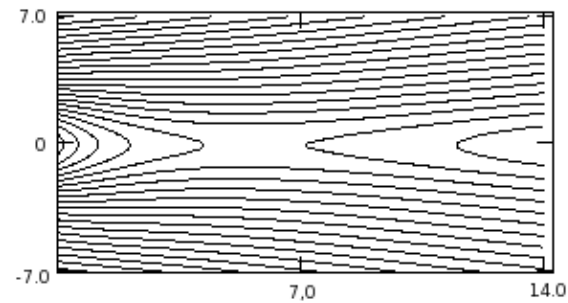}
  \caption{Evolution for the Magnetic field lines in the magnetic reconnection
  process, at $t=20/\Omega_1$}\label{tiempo20}
\end{figure}
Figures \ref{tiempo0}, \ref{tiempo3}, \ref{tiempo15} and \ref{tiempo20} show the evolution of the
magnetic field lines in the magnetic reconnection process. This
appears in a natural way, without the  {\it a priori} introduction of
any resistive region. The factor $\Omega_1$ is the ionic cyclotron frequency, $\Omega_1=q_1B_0/m_1$. This result tell us that the model can
actually simulate the magnetic reconnection. This simulation took 1h
in a Pentium IV PC of 2.8GHz, i.e. it is really fast.

\subsection{Reconnection rates}

To compute real reconnection rates we performed a similar simulation
to the one before, but with the actual ratio between electronic and
ionic masses ($m_1$$=1820m_0$). This choice bring us to take a shorter
time steps ($\delta t$$=$$3.76\times 10^{-5}$s) and smaller cells
($\delta x$$=$$15.95$km) in order to reproduce with accuracy the
electron moves. The LB array is $200$$\times$$100$ cells (larger in
direction x), for a total simulation region of $3190$km in x and
$1595$km in z. Since the region is smaller than before, $v(x)$$=$$1$
is a good approximation on the entire region. The simulation constants
are $L$$=$$1595$km \cite{n21} and $B_0$$=$$10.0$$nT$ \cite{n22}. The densities
in Eq.\eqref{tdensity} are $n_b$$=$$0$ and $n_c$$=$$10^{5}$$m^{-3}$ \cite{n21}, the
electronic temperature is chosen as $T_0$$=$$5.8$$MK$ and the ionic one as
$T_1$$=$$23.2$$MK$ \cite{n23}. All these are observational data. The
electronic mass is taken $m_0$$=$$9.11\times 10^{-31}$kg and the ionic mass is
$m_1$$=$$1.67\times 10^{-27}$kg. All other constants of our LB model take
their standard values in IS units. 

\begin{figure}
  \centering
  \includegraphics[scale=0.61]{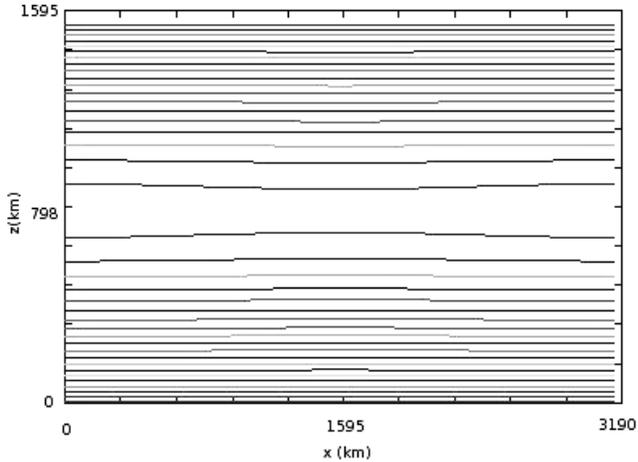}
  \caption{Magnetic field lines in the magnetic reconnection
  process at t=0 (initial conditions)}\label{tiempo01}
\end{figure}

\begin{figure}
  \centering
  \includegraphics[scale=0.61]{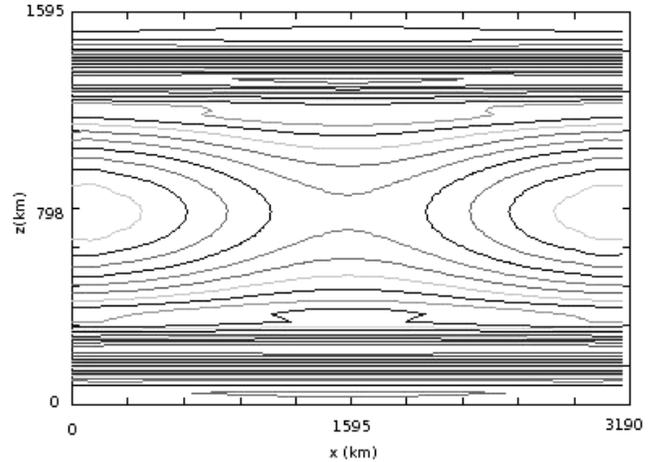}
  \caption{Evolution for the Magnetic field lines in the magnetic reconnection
  process, at $t=1.92ms$}\label{tiempo51}
\end{figure}

The initial configuration of the magnetic field is shown in figure
\ref{tiempo01} and the same field after $t$$=$$1.92 ms$ is shown in figure
\ref{tiempo51}. The reconnection rate we obtain from this simulation is
$R$$=$$0.109$, which is in good agreement with the experimental
observations around $R$$\sim$$0.1$ \cite{n28}. This simulation took just 
5 minutes in a Pentium IV PC of 2.8GHz.  
\section{Conclusion}
\label{conclusion}

In this paper we introduce a 3D lattice Boltzmann model for simulating
plasmas, which is able to simulate magnetic reconnection without any
previous assumption of a resistive region or an anomalous
resistivity. The model simulates the plasma as two fluids (one
electronic and one ionic) with an interaction term, and reproduces
in the continuous limit the equations of the two-fluids theory and,
therefore, the MHD-Hall equations.
This model can simulate either conducting and viscous fluids in the
incompressible limit or non-viscous compressible fluids, and
sucessfully reproduces both the Hartmann flow and the magnetic reconnection
in the magnetotail. The reconnection rate we obtain with this model is
$R$$=$$0.109$, which is in excellent agreement with observations.

Since this method includes both electric and magnetic fields, plus the
density and velocity fields for each fluid, it gives much more
information on the details of the plasma physics. Moreover, this
opens the door to much more sophisticated boundary conditions, like
conductive walls or electromagnetic waves in plasmas. This is an
advantage upon other magnetohydrodynamic LB models. Furthermore, it is
3D, so many interest phenomena can be investigated here. The model
does not require large computational resorces. It just takes between 5
minutes and 1h in a Pentium IV PC of 2.8GHz and uses around 100MB of
RAM.

The model introduces the forces at first order in time, but this is not
a problem for weak electromagnetic fields and low resistive
plasmas. If this is not the case, it is possible to modify the
charge/mass ratio, but this changes the MHD-Hall equations and slows
the evolution of the electromagnetic fields. Another way to increase
the numerical stability consists of modifying the model to reproduce
the two fluids in a different way: by defining density functions for the
sum, $f_i^{p(0)}+f_i^{p(1)}$, and the difference,
$\frac{q_0}{m_0}f_i^{p(0)}+\frac{q_1}{m_1}f_i^{p(1)}$ of the two
fluids. It is also possible to develop a LB model with 13 velocity
vectors for the fluids, as proposed by \cite{n26}. These are promisory
paths of future work.

Hereby we have introduced a 3D lattice Botzmann model that reproduces
the two-fluid theory and includes in a natural way many aspects of
interest in plasma physics, like electric fields and magnetic
reconnection. It has been shown in this work that this model can
actually be used to investigate real astrophysical problems. We hope
that this LB model will contribute to the study of plasma physics in
many interesting phenomena. 
   
\begin{acknowledgments}
The authors are thankful to Dominique d'Humi\`eres for his papers on the
method of lattice-Boltzmann.
\end{acknowledgments}

\appendix
\section{Chapman-Enskog Expansion} 
\label{ChapmanEnskog}
The Boltzmann equations for each fluid, Eq. \eqref{lbe0},
\eqref{lbe2} and \eqref{lbe3}, determine the system evolution. This
evolution rule gives in the continuum limit the macroscopic differential
equation that the system satisfies. This is known as the
Chapman-Enskog expansion. 
To develop it, we start by taking the Taylor expansion of these equations until
second order in spatial and temporal $(\delta \vec{x},\delta t)$ variables,  
\begin{eqnarray}{\label{lbee0}}
  \begin{aligned}
    &\vec{v}_i^p \cdot \vec{\nabla} f_{i}^{p(s)} \delta t +\frac{1}{2}
    \sum_{\alpha,\beta} \frac{\partial^2 f_{i}^{p(s)} }{\partial x_\alpha
      \partial x_\beta}(v_{i\alpha}^p v_{i\beta}^p)\delta t^2 \\ &+\frac{\partial f_{i}^{p(s)}}{\partial t} \delta
    t + \frac{\partial}{\partial t}\vec{v}_i^p \cdot \vec{\nabla} f_{i}^{p(s)} \delta
    t^2 \\&+\frac{1}{2}\frac{\partial^2 f_{i}^{p(s)}}{\partial t^2}\delta t^2= 
    -\frac{1}{\tau_s}(f_{i}^{p(s)}-f_{i}^{p(s)\rm eq}) \\&+\frac{\kappa_s \delta t}{20c^2}(\vec{v}_i^p \cdot
    \vec{F}^{(s)}) \quad ,
  \end{aligned}
\end{eqnarray}
\begin{eqnarray}{\label{lbee1}}
  \begin{aligned}
    &\vec{v}_i^p \cdot \vec{\nabla} f_{ij}^{p(2)} \delta t +\frac{1}{2}
    \sum_{\alpha,\beta} \frac{\partial^2 f_{ij}^{p(2)} }{\partial x_\alpha
      \partial x_\beta}(v_{i\alpha}^p v_{i\beta}^p)\delta t^2  \\& +\frac{\partial f_{ij}^{p(2)}}{\partial t} \delta
    t + \frac{\partial}{\partial t}\vec{v}_i^p \cdot \vec{\nabla} f_{ij}^{p(2)} \delta
    t^2 \\ &+ \frac{1}{2}\frac{\partial^2 f_{i}^{p(s)}}{\partial t^2}\delta t^2 = 
    -\frac{1}{\tau_2}(f_{ij}^{p(2)}-f_{ij}^{p(2)\rm eq})\\& -\frac{\kappa_2
      \mu_0 \delta t}{8}(\vec{e}_{ij}^p \cdot \vec{J'}) \quad , 
  \end{aligned}
\end{eqnarray}
\begin{eqnarray}{\label{lbee2}}
  \begin{aligned}
  \frac{\partial f_{0}^{(K)}}{\partial t}\delta t+\frac{1}{2}\frac{\partial^2
  f_{0}^{(K)}}{\partial t^2}\delta t^2=-\frac{1}{\tau_K}(f_{0}^{(K)}-f_{0}^{(K) \rm eq}) \quad .
  \end{aligned}
\end{eqnarray}

where $\alpha, \beta$$=$$x, y, z$ denotes the components in $x$, $y$ and $z$
 directions.

Next, we expand the distribution functions and the spatial and time
derivatives in a power series on a small parameter, $\epsilon$,
\begin{equation}
  f_{ij}^{p(2)}=f_{ij}^{p(2)(0)}+\epsilon f_{ij}^{p(2)(1)}+\epsilon^2
  f_{ij}^{p(2)(2)}+... \quad ,
\end{equation}
\begin{equation}
  f_{i}^{p(s)}=f_{i}^{p(s)(0)}+\epsilon f_{i}^{p(s)(1)}+\epsilon^2
  f_{i}^{p(s)(2)}+... \quad ,
\end{equation}
\begin{equation}
  \frac{\partial}{\partial t}=\epsilon \frac{\partial}{\partial
  t_1}+\epsilon^2 \frac{\partial}{\partial t_2}+... \quad ,
\end{equation}
\begin{equation}
    \frac{\partial}{\partial x_\alpha}=\epsilon \frac{\partial}{\partial
    x_{\alpha 1}}+... \quad .
\end{equation}
It is assumed that only the 0th order terms in $\epsilon$ of the
distribution functions contribute to the macroscopic variables. So,
for $n>0$ we have
\begin{subequations}{\label{nomacros}}
  \begin{equation}
    f_0^{s(n)} + \sum_{i,p} f_{i}^{p(s)(n)}=0 \quad ,
  \end{equation}
  \begin{equation}
    \sum_{i,p} f_{i}^{p(s)(n)} \vec{v}_i^p=0 \quad ,
  \end{equation}
  \begin{equation}
    \sum_{i,j,p} f_{ij}^{p(2)(n)} \vec{e}_{ij}^p=0 \quad ,
  \end{equation}
  \begin{equation}
    \sum_{i,j,p} f_{ij}^{p(2)(n)} \vec{b}_{ij}^p=0 \quad .
  \end{equation}
\end{subequations}

The external forces $\vec{F}^{(s)}$ and the current density $\vec{J'}$
are of order $\epsilon$ \cite{n14}, so we can write
$\vec{F}^{(s)}$$=$$\epsilon \vec{F}^{(s)}_1$ and $\vec{J'}$$=$$\epsilon
\vec{J'}_1$. Because $f_{i}^{p(s)\rm eq}$ and $f_{ij}^{p(2)\rm eq}$
are now functions of $\vec{F}^{(s)}$ and $\vec{J'}$, we need to develop
a Chapman-Enskog expansion of the equilibrium function, too: 
\begin{equation}
  f_{i}^{p(s)\rm eq}=f_{i}^{p(s)(0)\rm eq}+\epsilon f_{i}^{p(s)(1)\rm
  eq}+\epsilon^2 f_{i}^{p(s)(2)\rm eq} \quad ,
\end{equation}
\begin{equation}
  f_{ij}^{p(2)\rm eq}=f_{ij}^{p(2)(0)\rm eq}+\epsilon
  f_{ij}^{p(2)(1)\rm eq}+\epsilon^2 f_{ij}^{p(2)(2)\rm eq} \quad .
\end{equation}

Thus, by replacing these results into Eqs.\eqref{lbee0}, \eqref{lbee1}
and \eqref{lbee2}, we obtain at zeroth order of $\epsilon$
\begin{subequations}{\label{zeroth}}
\begin{equation}
  f_{i}^{p(s)(0)\rm eq}=f_{i}^{p(s)(0)} \quad ,
\end{equation}
\begin{equation}
  f_{0}^{(K)(0)\rm eq}=f_{0}^{(K)(0)} \quad ,
\end{equation}
\begin{equation}
  f_{ij}^{p(2)(0)\rm eq}=f_{ij}^{p(2)(0)} \quad .
\end{equation}
\end{subequations}

For the first order terms in $\epsilon$ of the distribution functions we obtain
\begin{subequations}{\label{first}}
\begin{eqnarray}{\label{firsta}}
  \begin{aligned}
    \vec{v}_i^p \cdot \vec{\nabla}_1 f_{i}^{p(s)(0)} \delta t &+\frac{\partial f_{i}^{p(s)(0)}}{\partial t_1} \delta
    t = \\&
    -\frac{1}{\tau_s}(f_{i}^{p(s)(1)}-f_{i}^{p(s)(1)\rm eq}) \\&+\frac{\kappa_s \delta
      t}{20c^2}(\vec{v}_i^p \cdot \vec{F}^{(s)}_1)   \quad ,
  \end{aligned}
\end{eqnarray}
\begin{eqnarray}{\label{firstb}}
  \begin{aligned}
    \vec{v}_i^p \cdot \vec{\nabla}_1 f_{ij}^{p(2)(0)} \delta t &+\frac{\partial f_{ij}^{p(2)(0)}}{\partial t_1} \delta
    t = \\& 
    -\frac{1}{\tau_2}(f_{ij}^{p(2)(1)}-f_{ij}^{p(2)(1)\rm eq}) \\&-\frac{\kappa_2 \mu_0 \delta t}{8}(\vec{e}_{ij}^p \cdot    \vec{J'}_1) \quad , 
  \end{aligned}
\end{eqnarray}
\begin{eqnarray}{\label{firstc}}
  \begin{aligned}
 \frac{\partial f_{0}^{(K)(0)}}{\partial t_1} \delta
  t =-\frac{1}{\tau_K}(f_{0}^{(K)(1)}-f_{0}^{(K)(1)\rm
  eq}) \quad ,
  \end{aligned}
\end{eqnarray}
\end{subequations}
and for the second order terms in $\epsilon$ we have
\begin{subequations}{\label{second}}
\begin{eqnarray}{\label{seconda}}
  \begin{aligned}
    &\biggl(1-\frac{1}{2\tau_s}\biggr)\biggl(\vec{v}_i^p \cdot \vec{\nabla}_1+\frac{\partial }{\partial
      t_1}\biggr)f_{i}^{p(s)(1)}\delta t \\&+ \frac{\partial f_{i}^{p(s)(0)}}{\partial
      t_2}\delta t + \frac{\delta t}{2\tau_s}\biggl(\vec{v}_i^p \cdot \vec{\nabla}_1+\frac{\partial }{\partial
      t_1}\biggr)f_{i}^{p(s)(1)\rm eq}\\&+\frac{\kappa_s \delta t}{40 c^2}\biggl(\vec{v}_i^p \cdot \vec{\nabla}_1+\frac{\partial }{\partial
      t_1}\biggr)(\vec{v}_i^p \cdot \vec{F}^{(s)}_1)=
    \\&-\frac{1}{\tau_s}(f_{i}^{p(s)(2)}-f_{i}^{p(s)(2)\rm eq}) \quad ,
  \end{aligned}
\end{eqnarray}
\begin{eqnarray}{\label{secondb}}
  \begin{aligned}
  &\biggl(1-\frac{1}{2\tau_2}\biggr)\biggl(\vec{v}_i^p \cdot \vec{\nabla}_1+\frac{\partial }{\partial
  t_1}\biggr)f_{ij}^{p(2)(1)}\delta t \\&+ \frac{\partial f_{ij}^{p(2)(0)}}{\partial
  t_2}\delta t + \frac{\delta t}{2\tau_2}\biggl(\vec{v}_i^p \cdot \vec{\nabla}_1+\frac{\partial }{\partial
  t_1}\biggr)f_{ij}^{p(2)(1)\rm eq} \\&+\frac{\mu_0 \kappa_2 \delta t}{16 c^2}\biggl(\vec{v}_i^p \cdot \vec{\nabla}_1+\frac{\partial }{\partial
  t_1}\biggr)(\vec{e}_{ij}^p \cdot \vec{J'}_1)=
  \\&-\frac{1}{\tau_2}(f_{ij}^{p(2)(2)}-f_{ij}^{p(2)(2)\rm eq}) \quad ,
  \end{aligned}
\end{eqnarray}
\begin{eqnarray}{\label{secondc}}
 \frac{\partial f_{0}^{(K)(0)}}{\partial t_1} \delta
  t = -\frac{1}{\tau_K}(f_{0}^{(K)(1)}-f_{0}^{(K)(1)\rm eq})   \quad .
\end{eqnarray}
\end{subequations}

The terms of order one and two for the equilibrium functions of the
fluids are obtained by replacing Eq. \eqref{expandV} into
Eq.\eqref{equilf}. That gives  
\begin{subequations}{\label{equilfe}}
\begin{eqnarray}
  \begin{aligned}
  f_{i}^{p(s)\rm eq}(\vec{x},t)&= \omega_i \rho_s \biggl[3 \xi_s
  \rho_s^{\gamma-1}+ \biggr. \\ & 3\biggl(\vec{v}_i^p \cdot
  \biggl(\vec{V}_{s}+\frac{ \epsilon \lambda_s \tau_s \delta t
  \vec{F}^{(s)}_1}{\rho_s}\biggr)\biggr)+ \\ &\frac{9}{4c^2}\biggl(\vec{v}_i^p \cdot
  \biggl(\vec{V}_{s}+\frac{ \epsilon \lambda_s \tau_s \delta t
  \vec{F}^{(s)}_1}{\rho_s}\biggr)\biggr)^2 \\
  & -\biggl.\frac{3}{2}\biggl(\vec{V}_{s}+\frac{ \epsilon \lambda_s
  \tau_s \delta t \vec{F}^{(s)}_1}{\rho_s}\biggr)^2 \biggr] \quad ,
  \end{aligned}
\end{eqnarray}
\begin{eqnarray}
  \begin{aligned}
  f_{0}^{p(s)\rm eq}(\vec{x},t)=& 6 \rho_s c^2\biggl(1-\frac{1}{4c^2}\biggl(4\xi_s
  \rho_s^{\gamma-1}+\\&\biggl(\vec{V}_{s}+\frac{ \epsilon
  \lambda_s \tau_s \delta t
  \vec{F}^{(s)}_1}{\rho_s}\biggr)^2\biggr)\biggr) \quad .
  \end{aligned}
\end{eqnarray}
\end{subequations}
From these equations we can obtain 
\begin{subequations}{\label{equilfes}}
\begin{eqnarray}
  \begin{aligned}
  f_{i}^{p(s)(0)\rm eq}(\vec{x},t)&=\omega_i \rho_s \biggl[3 \xi_s
  \rho_s^{\gamma-1}+3(\vec{v}_i^p \cdot
  \vec{V}_{s}) \biggr. \\ \biggl. &+\frac{9}{4c^2}(\vec{v}_i^p
  \cdot \vec{V}_{s})^2 - \frac{3}{2}(\vec{V}_{s})^2 \biggr] \quad ,
  \end{aligned}
\end{eqnarray}
\begin{eqnarray}
  \begin{aligned}
  f_{i}^{p(s)(1)\rm eq}(\vec{x},t)&=\omega_i \delta t\biggl[3\lambda_s \tau_s(\vec{v}_i^p
  \cdot\vec{F}^{(s)}_{1}) \biggr. \\ &+\frac{9 \lambda_s \tau_s}{2c^2}(\vec{v}_i^p
  \cdot \vec{V}_{s})(\vec{v}_i^p \cdot \vec{F}^{(s)}_{1}) \\ & - \biggl. 3\lambda_s
  \tau_s(\vec{V}_{s} \cdot \vec{F}^{(s)}_{1} ) \biggr] \quad ,
  \end{aligned}
\end{eqnarray}
\begin{eqnarray}
  \begin{aligned}
    f_{i}^{p(s)(2)\rm eq}(\vec{x},t)=\frac{\omega_i \delta t^2}{\rho_s}
    \biggl[\frac{9}{4c^2}\lambda^2_s \tau^2_s(\vec{v}_i^p \cdot \vec{F}^{(s)}_{1})^2 \biggr. \\ \biggl.-
    \frac{3}{2}\lambda^2_s \tau^2_s(\vec{F}^{(s)}_{1})^2 \biggr] \quad ,
  \end{aligned}
\end{eqnarray}
and
\begin{eqnarray}
  \begin{aligned}
  f_{0}^{p(s)(0)\rm eq}&(\vec{x},t)=\\&6 \rho_s
  c^2\biggl(1-\frac{1}{4c^2}(4\xi_s
  \rho_s^{\gamma-1}+(\vec{V}_{s})^2)\biggr) ,
  \end{aligned}
\end{eqnarray}
\begin{eqnarray}
  \begin{aligned}
  f_{0}^{p(s)(1)\rm eq}(\vec{x},t)=-6 \delta t c^2\biggl(\frac{\lambda_s \tau_s}{2c^2}(\vec{V}_{s}
  \cdot \vec{F}^{(s)}_1)\biggr) ,
  \end{aligned}
\end{eqnarray}
\begin{eqnarray}
  \begin{aligned}
  f_{0}^{p(s)(2)\rm eq}&(\vec{x},t)=\\&-\frac{6 \delta t^2
  c^2}{\rho_s}\biggl(\frac{\lambda^2_s \tau^2_s}{4c^2}(\vec{F}^{(s)}_1 \cdot
  \vec{F}^{(s)}_1)\biggr) ,
  \end{aligned}
\end{eqnarray}
\end{subequations}

The same process can be used to determine the terms of order one and
two for the equilibrium functions of the electromagnetic
fields. Replacing Eq. \eqref{expandE} into Eq. \eqref{equilc} and grouping, we have
\begin{subequations}{\label{equilces}}
\begin{eqnarray}
  f_{ij}^{p(2)(0)\rm eq}(\vec{x},t)=\frac{1}{8c^2}\vec{E} \cdot
  e_{ij}^{p}+\frac{1}{8}\vec{B} \cdot b_{ij}^{p} \quad ,
\end{eqnarray}
\begin{eqnarray}
  f_{ij}^{p(2)(1)\rm eq}(\vec{x},t)=-\frac{\epsilon \mu_0 \lambda_2
  \tau_2 \delta t}{8}\vec{J'}_1 \cdot e_{ij}^{p} \quad ,
\end{eqnarray}
\begin{eqnarray}
  f_{ij}^{p(2)(2)\rm eq}(\vec{x},t)=0 \quad .
\end{eqnarray}
\end{subequations}

Now, we are ready to determine the equation that the model satisfies
in the continuum limit. First, let us consider non-viscous compressible fluids,
that is $\tau_s$$=$$\frac{1}{2}$. By summing up
Eq. \eqref{firsta} over $i$ and $p$, and by taking into account
Eqs. \eqref{firstc}, \eqref{macros}, \eqref{equilfes} and
\eqref{nomacros}, we get 
\begin{eqnarray}{\label{firstsuma}}
  \vec{\nabla} \cdot (\rho_s \vec{V}_s)+\frac{\partial \rho_s}{\partial
  t_1}=0 \quad .
\end{eqnarray}
By summing up Eq. \eqref{seconda} in the same way, we obtain
\begin{eqnarray}{\label{secondsuma}}
  \vec{\nabla} \cdot \biggl(\frac{\lambda_s+\kappa_s}{2}\delta
  t\vec{F}^{(s)}_1\biggr)+\frac{\partial \rho_s}{\partial t_2}=0 \quad .
\end{eqnarray}
Now, we can add these two equations to obtain
\begin{eqnarray}{\label{firstsuma2}}
  \vec{\nabla} \cdot \biggl(\rho_s
  \vec{V}_s+\frac{\lambda_s+\kappa_s}{2}\delta
  t\vec{F}^{(s)}_1\biggr)+\frac{\partial \rho_s}{\partial t_1}=0 \quad .
\end{eqnarray}
Next, following Buick and Greated \cite{n14}, we do $\lambda_s$$=$$\frac{1}{2\tau_s}$,
$\kappa_s$$=$$\frac{2\tau_s-1}{2\tau_s}$ and, by taking into account
Eq. \eqref{expandV}, we arrive to the continuity equation
\begin{eqnarray}{\label{ecc}}
  \vec{\nabla} \cdot (\rho_s {\vec{V}'}_s)+\frac{\partial \rho_s}{\partial
  t}=0 \quad .
\end{eqnarray}

By multiplying Eq. \eqref{firsta} by $\vec{v}_i^p$ and summing up
over $i$ and $p$, we get
\begin{eqnarray}{\label{firstvsuma}}
  \frac{\partial}{\partial x_\beta}(\rho_s V_{s\alpha}
  V_{s\beta})+\frac{\partial (\xi_s \rho_s^{\gamma})}{\partial
  x_\alpha}+\frac{\partial (\rho_s V_{s\alpha})}{\partial
  t_1}=F_{1\alpha}^{(s)} .
\end{eqnarray}
In a similar way, by multiplying Eq. \eqref{seconda} by $\vec{v}_i^p$
and summing up over $i$ and $p$, we obtain
\begin{eqnarray}{\label{secondvsuma}}
  \frac{\partial (\rho_s V_{s\alpha})}{\partial t_2}+
  \frac{\delta t}{2}\frac{\partial}{\partial x_\beta}(F^{(s)}_{1\beta}
  V_{s\alpha}+F^{(s)}_{1\alpha}
  V_{s\beta}) \nonumber \\+\frac{\delta t}{2}\frac{\partial
  F_{1\alpha}^{(s)}}{\partial t_1}=0 \quad .
\end{eqnarray}
Now, we can add these two equations, and by replacing 
Eq. \eqref{expandV}, we get (up to second order in $\epsilon$)
\begin{eqnarray}{\label{NScp}}
  \frac{\partial (\rho_s V'_{s\alpha})}{\partial t}+\frac{\partial}{\partial
  x_\beta}(\rho_s {V'}_{s\alpha}{V'}_{s\beta})=-\frac{\partial
  P_s}{\partial x_\alpha}+F_{1\alpha}^{(s)} \quad .
\end{eqnarray}
This is the Navier-Stokes equation for non-viscous compressible fluids,
with state equation $P_s$$=$$\xi_s \rho_s^{\gamma}$.
In our model, the force $F_{\alpha}^{(s)}$ is taken at first order in
$\epsilon$. With this approximation, Eq.\eqref{force} gives
$F_{1\alpha}^{(s)}(\vec{V}_s)$$=$$F_{1\alpha}^{(s)}(\vec{V'}_s)$, and the
Navier-Stokes equation is
\begin{eqnarray}{\label{NSc}}
  \begin{aligned}
    \frac{\partial (\rho_s V'_{s\alpha})}{\partial t}&+\frac{\partial}{\partial
      x_\beta}(\rho_s {V'}_{s\alpha}{V'}_{s\beta})=\\&-\frac{\partial P_s}{\partial
      x_\alpha}+\biggl(\frac{q_s}{m_s}\rho_s(\vec{E}+\vec{V'}_s \times
    \vec{B}) \biggr. \\ \biggl. &-\nu
    \rho_s(\vec{V'}_s-\vec{V'}_{(s+1)mod 2})\biggr)_\alpha + F_{0\alpha} \quad .
  \end{aligned}
\end{eqnarray}
By replacing Eq.\eqref{ecc} into Eq.\eqref{NSc}, we arrive to the usual form
of the Navier-Stokes equation for a non-viscous compressible fluid \cite{n3}
\begin{eqnarray}{\label{NScV}}
  \begin{aligned}
    \rho_s \biggl (\frac{\partial \vec{V'}_s}{\partial t} &+ (\vec{V'}\cdot
      \vec{\nabla}) \vec{V'}_s \biggr)=\\&-\vec{\nabla} P_s+\frac{q_s}{m_s}\rho_s(\vec{E}+\vec{V'}_s \times
    \vec{B}) \\ &-\nu
    \rho_s(\vec{V'}_s-\vec{V'}_{(s+1)mod 2})+ \vec{F}_{0} \quad .
  \end{aligned}
\end{eqnarray}
Second, let us consider both fluids with viscosity ($\tau_s > 1/2$) in
the incompressible limit. By following the same procedure, we
arrive to the following momentum equation (up to second order in $\epsilon$):
\begin{eqnarray}{\label{NSi}}
  \begin{aligned}
    \frac{\partial (\rho_s V'_{s\alpha})}{\partial t}&+\frac{\partial}{\partial
      x_\beta}(\rho_s {V'}_{s\alpha}{V'}_{s\beta})= \\&-\frac{\partial P_s}{\partial
      x_\alpha}+\biggl(\frac{q_s}{m_s}\rho_s(\vec{E}+\vec{V'}_s \times
    \vec{B}) \biggr.\\ \biggl.&-\nu
    \rho_s(\vec{V'}_s-\vec{V'}_{(s+1)mod 2})\biggr)_\alpha\\& +\eta_s
    \rho_s \vec{\nabla}^2 V'_{s\alpha}+ F_{0\alpha} \quad ,
  \end{aligned}
\end{eqnarray}
where the kinematic viscosity is
$\eta_s$$=$$\frac{2}{3}(\tau_s-1/2)c^2 \delta t$. By following the same
procedure described above \cite{n3}, we arrive
\begin{eqnarray}{\label{NSiV}}
  \begin{aligned}
    \rho_s \biggl (\frac{\partial \vec{V'}_s}{\partial t} &+ (\vec{V'}\cdot
      \vec{\nabla}) \vec{V'}_s \biggr)=\\&-\vec{\nabla} P_s+\frac{q_s}{m_s}\rho_s(\vec{E}+\vec{V'}_s \times
    \vec{B}) \\ &-\nu
    \rho_s(\vec{V'}_s-\vec{V'}_{(s+1)mod 2})\\ &+ \vec{F}_{0} +\eta_s
    \rho_s \vec{\nabla}^2 \vec{V'}_{s}\quad .
  \end{aligned}
\end{eqnarray}
For the electromagnetic field, we take $\tau_2$$=$$1/2$,
$\lambda_2$$=$$1$ and $\kappa_2=0$. By summing up Eqs. \eqref{firstb} and \eqref{secondb}
on $i$, $j$ and $p$, we do not get any information about the
fields. Thus, let us multiply these equations by $\vec{e}_{ij}^p$ before
summing up. So, we obtain
\begin{eqnarray}{\label{Max10}}
  \frac{\partial \vec{E}}{\partial t_1}-c^2\vec{\nabla} \times \vec{B}=-\mu_0
  c^2 \vec{J'}_1 \quad ,
\end{eqnarray}
and
\begin{eqnarray}{\label{Max11}}
  \frac{\partial \vec{E}}{\partial t_2}-\frac{\mu_0 c^2 \delta
  t}{2}\frac{\partial \vec{J'}_1}{\partial t_1}=0 \quad .
\end{eqnarray}
If we add these two equations, and because of Eq. \eqref{expandE}, we get
the first Maxwell equation,
\begin{eqnarray}{\label{Max12}}
  \frac{\partial \vec{E'}}{\partial t}-c^2\vec{\nabla} \times \vec{B}=-\mu_0
  c^2 \vec{J'} \quad .
\end{eqnarray}
Similarly, multiplying Eqs. \eqref{firstb} and \eqref{secondb} by $\vec{b}_{ij}^p$
and summing up on $i$, $j$ and $p$, we obtain
\begin{eqnarray}{\label{Max20}}
  \frac{\partial \vec{B}}{\partial t_1}+\vec{\nabla} \times \vec{E}=0 \quad ,
\end{eqnarray}
and
\begin{eqnarray}{\label{Max21}}
  \frac{\partial \vec{B}}{\partial t_2}-\frac{1}{2}\vec{\nabla} \times
  (\mu_0 c^2 \delta t \vec{J'}_1)=0 \quad .
\end{eqnarray}
If we add these two equations, we obtain the second Maxwell equation,
\begin{eqnarray}{\label{Max22}}
  \frac{\partial \vec{B}}{\partial t}+\vec{\nabla} \times \vec{E'}=0 \quad .
\end{eqnarray}
The other two Maxwell equations can be obtained from the Eqs.\eqref{Max12} and
\eqref{Max22} as follows \cite{n3}. If one applies the divergence to these
equations we obtain 
\begin{eqnarray}{\label{Max30}}
  \frac{\partial (\vec{\nabla} \cdot \vec{E'})}{\partial t}=-\mu_0
  c^2 \vec{\nabla} \cdot \vec{J'} \quad ,
\end{eqnarray}
\begin{eqnarray}{\label{Max40}}
  \frac{\partial (\vec{\nabla} \cdot \vec{B})}{\partial t}=0 \quad .
\end{eqnarray}
Now, we replace the Eq.\eqref{currentequil} in the Eq.\eqref{Max30} to get
\begin{eqnarray}{\label{Max31}}
  \begin{aligned}
    \frac{\partial (\vec{\nabla} \cdot \vec{E'})}{\partial t}&=\\&-\mu_0
    c^2 \biggl(\frac{q_0}{m_0} \vec{\nabla} \cdot (\rho_0\vec{V'}_0)+\frac{q_1}{m_1}
    \vec{\nabla} \cdot (\rho_1 \vec{V'}_1) \biggr) ,
  \end{aligned}
\end{eqnarray}
and because of the two fluids satisfy the continuity equations \eqref{ecc}, we
obtain
\begin{eqnarray}{\label{Max32}}
  \begin{aligned}
    \frac{\partial (\vec{\nabla} \cdot \vec{E'})}{\partial t}=\mu_0
    c^2 \biggl(\frac{q_0}{m_0} \frac{\partial \rho_0}{\partial t}+\frac{q_1}{m_1}
    \frac{\partial \rho_1}{\partial t} \biggr) \quad .
  \end{aligned}
\end{eqnarray}
By taking into account the Eq. \eqref{macros}, we finally get
\begin{eqnarray}{\label{Max33}}
  \begin{aligned}
    \frac{\partial (\vec{\nabla} \cdot \vec{E'}-\mu_0c^2 \rho_c )}{\partial t}=0 \quad .
  \end{aligned}
\end{eqnarray}
Thus, if the initial conditions for the electromagnetic fields satisfy the
Maxwell equations
\begin{eqnarray}{\label{Max41}}
  \vec{\nabla} \cdot \vec{B}=0 \quad .
\end{eqnarray}
\begin{eqnarray}{\label{Max34}}
  \begin{aligned}
    \vec{\nabla} \cdot \vec{E'}=\mu_0c^2 \rho_c =  \frac{\rho_c}{\epsilon_0} \quad .
  \end{aligned}
\end{eqnarray}
this equations will be recovered for all times.

Summarizing, 
the state equation and Eqs. \eqref{ecc}, \eqref{NSc} determine
the behavior of a non-viscous compressible plasma. If we use 
Eq.\eqref{NSi} instead of  Eq.\eqref{NSc}, the model reproduces the behavior of
an incompressible plasma with viscosity. Eqs. \eqref{Max12},
\eqref{Max22} \eqref{Max41} and \eqref{Max34} determine the evolution of the electromagnetic
fields. These are the equations of the two-fluids theory \cite{n3}, and 
this completes the proof.  
 
\bibliography{tesis}

\end{document}